\documentclass[conference]{IEEEtran}
\IEEEoverridecommandlockouts

\usepackage{cite}
\usepackage{amsmath,amssymb,amsfonts}
\usepackage{algorithmic}
\usepackage{graphicx}
\usepackage{textcomp}
\usepackage{xcolor}
\usepackage{multirow}
\def\BibTeX{{\rm B\kern-.05em{\sc i\kern-.025em b}\kern-.08em
    T\kern-.1667em\lower.7ex\hbox{E}\kern-.125emX}}
\begin{document}

\title{X-CrossNet: A complex spectral mapping approach to target speaker extraction with cross attention speaker embedding fusion\\
}

\author{\IEEEauthorblockN{1\textsuperscript{st} Chang Sun}
\IEEEauthorblockA{\textit{AI Innovation Center} \\
\textit{FRI of the Ministry of Public Security of PRC}\\
Beijng, China \\
sunch@cnu.edu.cn}
\and
\IEEEauthorblockN{2\textsuperscript{nd} Bo Qin}
\IEEEauthorblockA{\textit{AI Innovation Center} \\
\textit{FRI of the Ministry of Public Security of PRC}\\
Beijng, China \\
bq@fri.com}
}

\maketitle

\begin{abstract}
Target speaker extraction (TSE) is a technique for isolating a target speaker's voice from mixed speech using auxiliary features associated with the target speaker. It is another attempt at addressing the cocktail party problem and is generally considered to have more practical application prospects than traditional speech separation methods. Although academic research in this area has achieved high performance and evaluation scores on public datasets, most models exhibit significantly reduced performance in real-world noisy or reverberant conditions. To address this limitation, we propose a novel TSE model, X-CrossNet, which leverages CrossNet as its backbone. CrossNet is a speech separation network specifically optimized for challenging noisy and reverberant environments, achieving state-of-the-art performance in tasks such as speaker separation under these conditions.
Additionally, to enhance the network's ability to capture and utilize auxiliary features of the target speaker, we integrate a Cross-Attention mechanism into the global multi-head self-attention (GMHSA) module within each CrossNet block. This facilitates more effective integration of target speaker features with mixed speech features. Experimental results show that our method performs superior separation on the WSJ0-2mix and WHAMR! datasets, demonstrating strong robustness and stability.
\end{abstract}


\section{Introduction}
In most real-world applications of voice interaction systems, the speech signals captured by microphones often include multiple speakers, are contaminated by noise, and may also be affected by reverberation. These factors pose significant challenges for tasks such as speech recognition. To address this, speech enhancement modules are commonly employed as a preprocessing step to improve the quality of speech signals for tasks like speech recognition. Speech separation is one approach to speech enhancement and includes various research directions, such as blind source separation (BSS) and target speaker extraction.

Blind source separation aims to separate all speaker channels from mixed speech signals. While BSS performs well on benchmark datasets like WSJ0-2mix \cite{wsj02mix}, its application in real-world scenarios is challenging due to the unpredictable number of speakers. Target speaker extraction, which can be considered either an improvement on or a subfield of BSS, offers a more feasible approach. Unlike BSS, which relies solely on mixed speech as input, TSE also uses a reference speech segment from the target speaker. By leveraging the features of the reference speech, TSE identifies and extracts the target speaker while ignoring other speakers and background noise. From a technical perspective, TSE simplifies the complexity of BSS, and from an application standpoint, it is better suited to practical scenarios.

As mentioned earlier, target speaker extraction builds upon blind source separation. Recent advancements in TSE research have primarily focused on improving state-of-the-art BSS methods by leveraging their established architectures as mature separation modules. For instance, X-TasNet \cite{xtasnet} is derived from Conv-TasNet \cite{tasnet}, X-Sepformer \cite{xsepformer} builds upon SepFormer \cite{sepformer}, and X-TF-GridNet \cite{xtfgridnet} uses TF-GridNet \cite{tfgridnet} as its backbone. Since these backbone BSS methods have demonstrated effectiveness in speech separation tasks, TSE models concentrate on incorporating reference speech feature extraction and integration into the backbone network. Key research questions in TSE involve the effective extraction of reference speech features and the optimal utilization of the reference speaker's information to enhance the separation network’s ability to filter or guide post-separation feature selection.

Early studies on reference speech feature extraction relied on pre-trained modules from speaker recognition works \cite{speakerbeam} \cite{voicefilter}, which did not participate in the training of the TSE model. In contrast, modern approaches integrate speech feature extraction modules into the joint training process with the backbone network, enabling the features to be more specifically adapted to the TSE task. Regarding integrating reference speech features, earlier methods simply concatenated mixed speech features with reference speech features after encoding. More recent techniques have adopted sophisticated integration strategies, such as multiplication \cite{multiplication}, scaling adaptation \cite{wavesplit}, and attention-based mechanisms \cite{abtse1} \cite{abtse2} \cite{abtse3}, or have introduced specifically designed modules like adaptive embedding fusion (AEA).

Despite significant advancements in target speaker extraction (TSE) technology in recent years, its application in real-world scenarios remains limited due to the challenges posed by noise and reverberation in mixed speech signals. CrossNet \cite{crossnet}, a blind source separation (BSS) model, is specifically designed to address these issues in noisy and reverberant environments. It utilizes a time-frequency domain approach with a structure consisting of a global multi-head self-attention (GMHSA) module, a cross-band module, a narrow-band module, and a decoder layer. The GMHSA module captures global correlations within the input signal, the cross-band module captures cross-band correlations, and the narrow-band module extracts information from neighboring frequency bins. Finally, the decoder layer reconstructs the separated features into a time-frequency representation. CrossNet has demonstrated state-of-the-art performance in tasks such as reverberant and noisy reverberant speaker separation. Therefore, CrossNet was selected as the backbone network to improve the robustness and stability of TSE tasks in challenging acoustic conditions.

In this paper, we propose a novel TSE model based on CrossNet, called X-CrossNet. This study has two primary objectives: (i) to extend CrossNet for TSE tasks with minimal parameter additions and without introducing new integration structures, and (ii) to enhance the robustness and stability of TSE tasks in noisy and reverberant environments, thereby advancing their practical applications.

The rest of the paper is organized as follows. Section \ref{tsep} describes the single-channel target speaker extraction problem in the time-frequency (T-F) domain. The detailed description of X-CrossNet is given in Section \ref{model}. Section \ref{exp} presents the experimental setup and the evaluation and comparison results. The last section is a summary of this article.

\begin{figure}
\centerline{\includegraphics[width=2.8in]{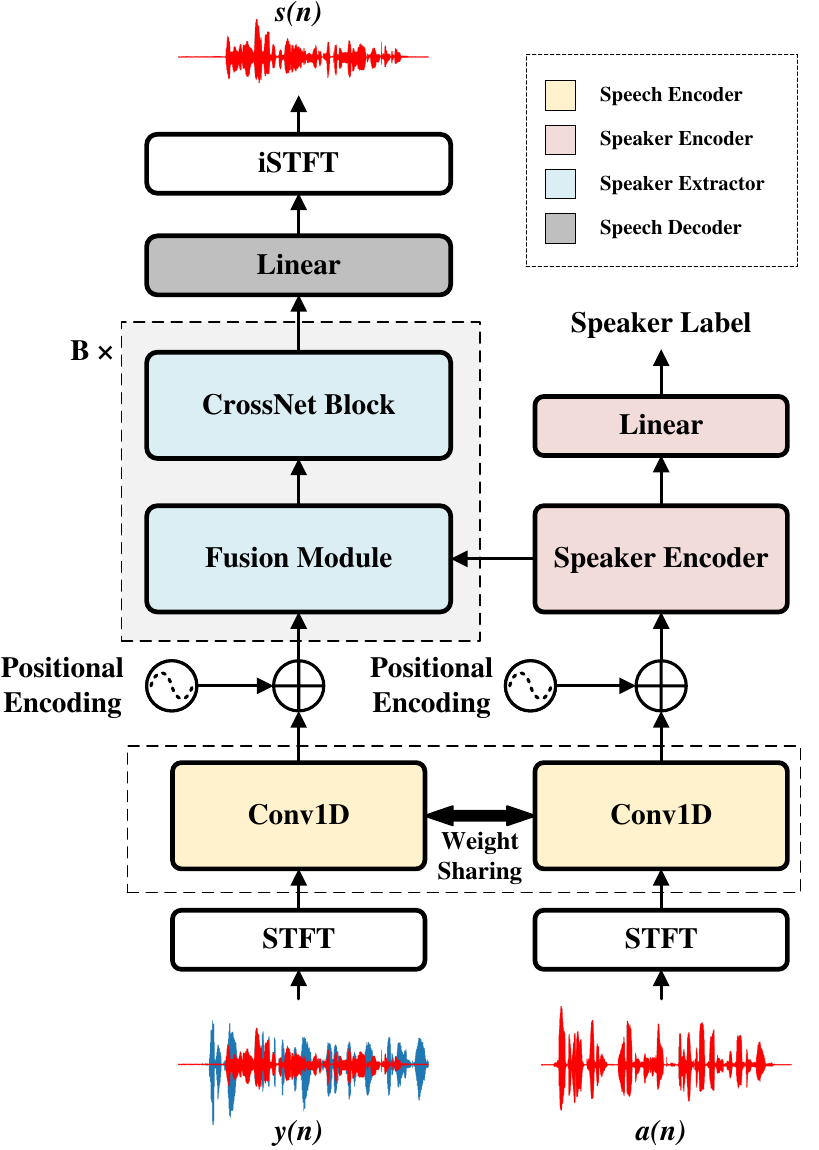}}
\caption{Schematic diagram of the proposed X-CrossNet, encompassing a speech encoder, a speaker encoder, a speaker extractor, and a speech decoder. The \(\mathbf{y}(n)\) is a mixture input signal, the \(\mathbf{a}(n)\) is the auxiliary signal of the target speaker, and the \(\mathbf{s}(n)\) is the target speech. \label{fig:fig1}}
\end{figure}

\begin{figure*}[h]
\centerline{\includegraphics[width=2\columnwidth]{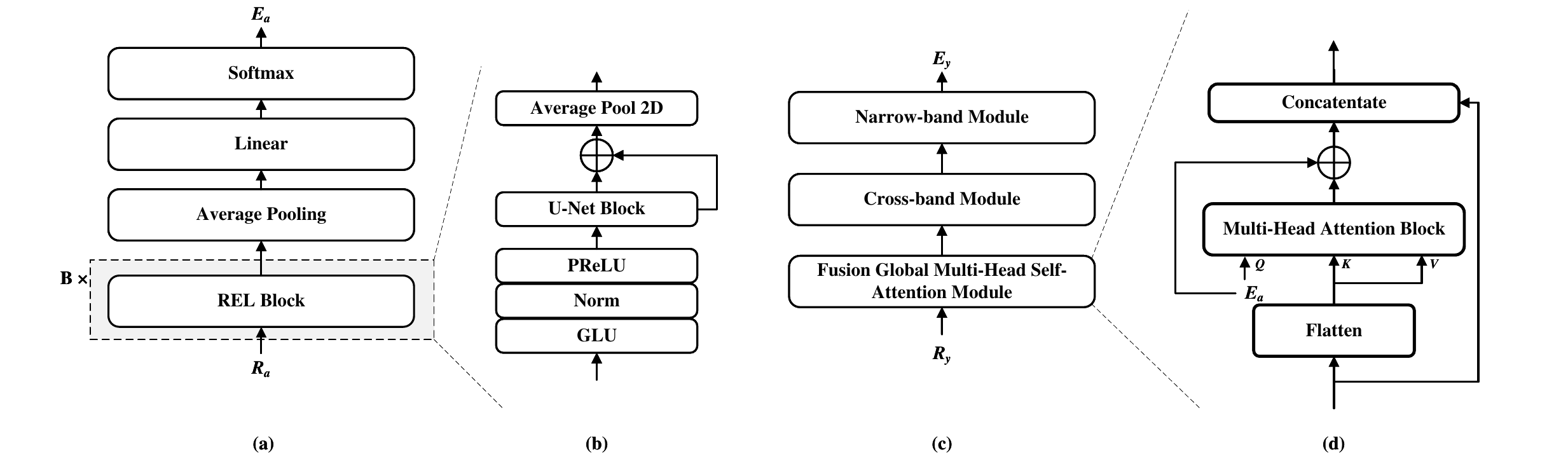}}
\caption{X-CrossNet building blocks. (a) Speaker Encoder module. (b) U-Net-based REL block. (c) CrossNet block. (d) Fusion global multi-head self-attention module.}
\label{fig:fig2}
\end{figure*}

\section{Target Speaker Extraction Problem}\label{tsep}
In the time domain, the mixed speech \(\mathbf{y}(n) \in \mathbb{R}\) of \(C\) speakers receiving noise and reverberation pollution can be expressed as: 
\begin{eqnarray}
\mathbf{y}\left ( n\right )=\displaystyle\sum_{c=1}^{C}\left ({\mathrm{s}}_{c}\left ( n\right )+{\mathrm{h}}_{c}\left ( n\right )\right )+\mathrm{v}\left ( n\right ), 
\end{eqnarray}
where \(n\in \left [ 1,...,N\right ]\) is the time index, and \(c\) is the speaker index. \({\mathrm{s}}_{c} (n)\), \({\mathrm{h}}_{c}(n)\), and \(\mathrm{v}(n)\) respectively represent recorded speech, reverberation, and noise. Correspondingly, in the time-frequency domain, the mixed speech collected by a single microphone contains \(C\) speakers, noise, and reverberation noise, which can be modeled as:
\begin{eqnarray}
\mathbf{Y}\left ( t,f\right )=\displaystyle\sum_{c=1}^{C}\left ({\mathbf{S}}_{c}\left ( t, f\right )+{\mathbf{H}}_{c}\left ( t, f\right )\right )+\mathbf{V}\left ( t, f\right ),
\end{eqnarray}
where \(t \in \left [ 1,...,T\right ]\) indices time frames and \(f \in \left [ 1,...,F\right ]\) frequency bins. For the TSE task, for the target speaker \(c\), \(\mathbf{A}_{c}\left ( t, f\right )\) represents the time-frequency domain representation of an enrollment utterance \({\mathrm{a}}_{c}(n)\), and the TSE task can be represented as: 
\begin{eqnarray}
{\mathbf{\hat{S}}}_{c}\left ( t,f\right )=f\left ( {\mathbf{Y}}\left ( t,f\right ), {\mathbf{A}}_{c}\left ( t,f\right ), \varPhi \right ),
\end{eqnarray}
where the \({\mathbf{\hat{S}}}_{c}(t,f)\) represents the target speech to be extracted and the \(\varPhi\) is the trainable parameter of the model. For a complex network, by predicting the real part \({\mathbf{\hat{S}}}_{{c}_{r}}\) and imaginary part \({\mathbf{\hat{S}}}_{{c}_{i}}\) of the direct path signal of the target speaker separately, the above equation can be expressed as:
\begin{eqnarray}
\left ({\mathbf{\hat{S}}}_{{c}_{r}}, {\mathbf{\hat{S}}}_{{c}_{i}}\right )=f\left ( \left ( {\mathbf{Y}}_{r}, {\mathbf{Y}}_{i}\right ), \left ( {\mathbf{A}}_{{c}_{r}}, {\mathbf{A}}_{{c}_{i}}\right ), \varPhi \right ),
\end{eqnarray}
where \(\varPhi\) is the trainable parameters of the model, and \({\mathrm{A}}_{{c}_{r}}\) and \({\mathrm{A}}_{{c}_{i}}\) are the real and image parts of the enrollment utterance. Finally, the direct path signal of target speaker \({\mathrm{\hat{s}}}_{c}(n)\) was restored through inverse short-time Fourier transform (iSTFT).

\section{X-CrossNet}\label{model}
This section introduces the model structure of X-CrossNet. Figure \ref{fig:fig1} shows the structure of the model, which is divided into four main parts: speech encoder, speech decoder, speaker encoder, and speaker extractor. The input mixed speech \(\mathbf{y}(n)\) and enrollment correction \(\mathbf{a}(n)\) are first transformed into their frequency-domain representations \({\mathbf{Y}}(t,f)\) and \({\mathbf{A}}(t,f)\) using the short-time fourier transform (STFT). These two signals are processed by the speech encoder, which shares weights between them, to generate high-dimensional feature representations \({\mathbf{R}}_{y}\) and \({\mathbf{R}}_{a}\). Next, \({\mathbf{A}}(t,f)\) is processed by the speaker encoder to extract the speaker embedding \({\mathbf{E}}_{a}\). The speaker extractor combines \({\mathbf{E}}_{a}\) and \({\mathbf{R}}_{y}\) to produce the separated speech representation \({\mathbf{E}}_{y}\). Finally, \({\mathbf{E}}_{y}\) is passed through the speech decoder to generate \({\mathbf{\hat{S}}}_{{c}_{r}}\) and \({\mathbf{\hat{S}}}_{{c}_{i}}\), which are then converted back into the extracted speech signal \({\mathrm{s}}(n)\) via the inverse iSTFT.

\subsection{Speech Encoder}
There is a weight-sharing speech encoder for both \(\mathbf{y}(n)\) and \(\mathbf{a}(n)\). The structure of the speech encoder is a Conv1D layer with a kernel size of \(k\) and a stride of 1. The speech encoder converts the scale of the inputs from \([2, \mathbf{F}, \mathbf{T}]\) to \([\mathbf{H}, \mathbf{F}, \mathbf{T}]\), in which 2 denotes the real and imaginary parts, and \(\mathbf{F}\) is the number of frequency bins, \(\mathbf{T}\) is the number of frames, and the \(\mathbf{H}\) is the number of hidden channels.

\subsection{Speaker Encoder}
The speaker encoder is designed to extract speaker-specific features from the enrollment utterance \(\mathbf{a}(n)\) to guide the speaker extractor module in performing speech separation. This component is based on the speaker encoder introduced in \cite{xtfgridnet} and adopts a structure centered on a \(\mathrm{B}\)-layer REL Blocks, as shown in Figure \ref{fig:fig2} (a). The REL Block is constructed based on a lightweight \({\mathrm{U}}^{2}\)-Net architecture, as depicted in Figure \ref{fig:fig2} (b), and primarily consists of a gated linear unit (GLU) block and a \({\mathrm{U}}\)-Net block. The speaker encoder takes the output \({\mathbf{R}}_{a}\) from the speech encoder as an input and produces the speaker feature \({\mathbf{E}}_{a}\). In addition to being fed into the speaker extractor to assist in speaker extraction, \({\mathbf{E}}_{a}\) is passed through a linear layer for speaker classification and compared with speaker labels to constrain the optimization process of the speaker encoder.

\subsection{Speaker Extractor}
The backbone network of the speaker extractor is derived from \cite{crossnet}. Similarly, we have introduced a random chunk positional encoding (RCPE) module after the speech encoder to generalize the model to longer sequences. As shown in Figure \ref{fig:fig2} (c), CrossNet mainly consists of \(\mathrm{B}\)-layer CrossNet Blocks. Each block contains three main modules, global multi-head self-attention module (GMHSA), cross-band module, and narrow-band module. To fuse with speaker features \({\mathbf{E}}_{a}\), we have improved the GMHSA of each block layer by adding a cross-attention-based fusion structure as shown in Figure \ref{fig:fig2} (d). The cross-band module consists of two frequency convolutional modules and a full-band linear module, where the frequency convolutional module contains a layer normalization (LN) layer, A grouped convolution layer along the frequency axis, and a parametric rectified linear unit (PReLU) activation function. The full-band linear module includes a linear layer, a sigmoid-weighted linear unit (SiLU), and a set of linear layers along the frequency axis The narrow-band module is a modified version of the Conformer \cite{conformer} convolutional block, which includes an LN layer, a linear layer, a SiLU activation, a time-convolutional (T-Conv) layer and a final linear layer.

\subsection{Speech Decoder}
The speech decoder mainly consists of a linear layer, which converts \({\mathbf{E}}_{y}\) into two channel outputs, simulating the real and imaginary parts of the output signal spectrum \({\mathbf{\hat{S}}}\left ( t,f\right )\).

\subsection{Loss functions}
A combination loss function is used in this paper to simultaneously constrain the training process of four modules, mainly including three parts: magnitude loss, scale-invariant signal-to-distortion ratio (SI-SDR) \cite{sisdr} loss, and cross-entropy (CE) loss. The loss functions are defined below:

\begin{eqnarray}
\mathcal{L}={\mathcal{L}}_{\mathrm{Mag}}+{\mathcal{L}}_{\mathrm{SI-SDR}}+{\mathcal{L}}_{\mathrm{CE}}
\end{eqnarray}

\begin{eqnarray}
{\mathcal{L}}_{\mathrm{Mag}}=\frac{{\left \| \left | \mathrm{STFT}\left ( \hat{{s}}\right )\right |-\left | \mathrm{STFT}\left ( {s}\right ) \right |\right \|}_{1}}{\left \|\left | \mathrm{STFT}\left ( {s}\right ) \right |\right \|_{1}}
\end{eqnarray}

\begin{eqnarray}
{\mathcal{L}}_{\mathrm{SI-SDR}}=-10\log_{10}{\frac{{\left \| {\alpha }{s}_{1}\right \|}_{2}^{2}}{{\left \| \hat{{s}} - {\alpha }{s}_{1}\right \|}_{2}^{2}}},  {\alpha }=\frac{\left \langle \hat{s},{s}_{1} \right \rangle}{\left \langle {s}_{1},{s}_{1} \right \rangle}
\end{eqnarray}

\begin{eqnarray}
{\mathcal{L}}_{\mathrm{CE}}=-\displaystyle\sum_{n=1}^{{N}_{s}}{p}_{n}\log\left ( {{\hat{p}}_{n}}\right )
\end{eqnarray}

The \({\mathcal{L}}_{\mathrm{CE}}\) is mainly used for speaker classification where the \({N}_{s}\) is the number of speakers. The setting of the magnitude loss refers to \cite{tfgridnet}, mainly used for the magnitude of the target signal in the STFT domain. In \({\mathcal{L}}_{\mathrm{Mag}}\), \({\left \| \cdot \right \|}_{1}\) is the \({L}_{1}\) norm, and \({\left | \cdot \right |}_{1}\) is the magnitude operator. In SI-SDR loss,  \({\left \| \cdot \right \|}_{2}^{2}\) is the \({L}_{2}\) norm, \(\alpha\) denotes the optimal scale factor and \(\left \langle \cdot , \cdot \right \rangle\) denotes the linear product.

\section{Experimental setups}\label{exp}
\subsection{Datasets}
To evaluate the performance of X-CrossNet and robustness in noisy and reverberant scenarios, we conducted two sets of experiments using datasets with and without noise. First, for single-channel speaker separation in anechoic conditions, following the methodology outlined in \cite{spexplus}\footnote{https://github.com/xuchenglin28/speaker\_extraction\_SpEx/tree/master/\\data/wsj0\_2mix}, we simulated a two-speaker database, WSJ0-2mix-extr, with a sampling rate of 8kHz. The dataset comprises 119 speakers and 28,000 utterances, divided as follows: the training set contains 20,000 utterances, and the development set includes 5,000 utterances, both drawn from 101 speakers in the "\(si\_tr\_s\)" corpus of the WSJ0 dataset. The test set consists of 3,000 utterances from 18 distinct speakers sourced from the "\(si\_dt\_05\)" and "\(si\_et\_05\)" subsets of the WSJ0 dataset. For each utterance, two speakers are randomly selected from a predefined pool, and one audio sample from each speaker is chosen and mixed at a signal-to-noise ratio (SNR) of 0–5 dB. Additionally, one audio sample from each speaker is used as reference speech to evaluate scenarios where either speaker serves as the target speaker.

We employ WHAMR! \cite{whamr} dataset to evaluate the robustness of the model in noisy and reverberant conditions. WHAMR! dataset is a further expanded vision of WHAM! \cite{wham} dataset by introducing artificial reverberation effects and environmental noise. WHAM! dataset extends WSJ0-2mix by incorporating various types of real-world background noise, including sounds recorded in environments such as restaurants, cafes, bars, and parks. WHAMR! dataset includes 20,000, 5,000, and 3,000 mixed utterances for training, validation, and testing, respectively.

\renewcommand\arraystretch{1.2}
\begin{table}[htbp]
\caption{Comparison results on the WSJ0-2mix dataset.}
\begin{center}
\begin{tabular}{|c|c|c|c|c|}
\hline
\textbf{Methods} & \textbf{Domain} & \textbf{Params (M)}& \textbf{SI-SDRi}& \textbf{SDRi} \\
\hline
TseNet \cite{tsenet} & T & 23.6& 12.2 & 12.6 \\
SpEx \cite{spex} & T & 10.8& 15.8 & 16.3 \\
SpEx+& T & 11.1& 16.9 & 17.2 \\
SEF-Net \cite{sefnet} &T& 27& 17.2 & 17.6 \\
X-SepFormer& T & 26.66 & 18.9 & 19.5 \\
\hline
X-TF-GridNet &T-F& 7.8& 19.7 & 20.4 \\
\hline
X-CrossNet&T-F& \textbf{5.1} & \textbf{19.9} & \textbf{20.5} \\
\hline
\end{tabular}
\vspace{-\baselineskip}
\label{tab1}
\end{center}
\end{table}

\renewcommand\arraystretch{1.2}
\begin{table}[htbp]
\caption{Comparison results on the WHAMR! dataset.}
\begin{center}
\begin{tabular}{|c|c|c|c|c|}
\hline
\textbf{Cond.} & \textbf{Methods} & Ref. length(s) & \textbf{SI-SDRi}& \textbf{SDRi} \\ \hline
\multirow{5}{*}{Noise + Rever.} & SpEx & 7.3 (Avg.)& 10.3 & 9.5 \\
 & SpEx+ & 7.3 (Avg.)& 10.8 & 10.1 \\
 & SpEx++ & 7.3 (Avg.)& 11.6 & 10.7 \\ 
 & X-TF-GridNet & 7.3 (Avg.)& 14.6 & 13.8 \\ \cline{2-5}
 & X-CrossNet & 7.3 (Avg.)& \textbf{14.6} & \textbf{14.1} \\
\hline
\end{tabular}
\vspace{-\baselineskip}
\label{tab2}
\end{center}
\end{table}

\subsection{Configurations}
We set the hyper-parameters of the model based on the work of \cite{tfgridnet} and \cite{spatialnet}. We use 4-second utterances input for both 2 datasets and use a 128 samples frame length Hanning window STFT to process the input signal. For the model architecture, we set hidden channel sizes to 96, and the kernel size of the Conv1D encoder is 5. We employ 12 layers of CrossNet Blocks and set the number of attention heads to 4 for the GMHSA module. During the training phase, we use an AdamW \cite{adamw} optimizer and a WarmupCosineScheduler scheduler, the maximum learning rate is 1e-3. We set the warmup epoch to 10 and the max epoch to 110. The model was trained on 2 NVIDIA RTX 4090 GPUs and we employ the half-precision (mixed-16) training strategy to reduce the GPU memory cost.

\subsection{Evaluation results}
We use signal-to-distortion ratio improvement (SDRi) \cite{sdri} and SI-SDR improvement (SI-SDRi) \cite{sisdr} as evaluation metrics, where higher values indicate better model performance. First, we compared the performance of X-CrossNet with state-of-the-art (SOTA) models on the WSJ0-2mix dataset. The results are presented in Table \ref{tab1}. The second column of the table specifies the input type for each method: X-TF-GridNet and X-CrossNet operate in the time-frequency domain, while the other methods function in the time domain. As shown in the table, X-CrossNet outperforms other methods on both SDRi and SI-SDRi metrics and achieves this with the lowest number of parameters. For experiments on the WHAMR! dataset, which includes noise and reverberation conditions, results are presented in Table \ref{tab2}. The third column of the table shows the length of the reference speech used, with 7.3 seconds selected for comparison. The results demonstrate that the proposed method achieves superior performance compared to existing approaches, confirming that X-CrossNet exhibits strong robustness and stability in noisy and reverberant scenarios.

\section{Conclusion}
In this paper, we propose a time-frequency domain target speaker extraction method named X-CrossNet, which uses CrossNet as the backbone network and is designed to perform effectively in noisy and reverberant environments. We introduce a REL block based speaker encoder to extract accurate speaker representations and enhance the GMHSA module in CrossNet by incorporating a fusion structure for speaker embeddings. Experimental results demonstrate that X-CrossNet achieves state-of-the-art performance on conventional test sets while maintaining significant robustness and stability in noisy and reverberant scenarios.

\end{document}